\documentclass{PoS}
\title{Observations of the Crab Nebula with H.E.S.S. Phase II}

\ShortTitle{Observations of the Crab Nebula with H.E.S.S. Phase II}

\author{\speaker{M. Holler}$^{a}$, D. Berge$^{b}$, C. van Eldik$^{c}$, J.-P. Lenain$^{d}$, V. Marandon$^{e}$, T. Murach$^{f}$, M. de Naurois$^{a}$, R.D. Parsons$^{e}$, H. Prokoph$^{g}$, D. Zaborov$^{a}$\thanks{On leave from Institute for Theoretical and Experimental Physics, B. Cheremushkinskaya 25, Moscow, 117218 Russia}~, for the H.E.S.S. collaboration\\
        ${}^{a}$Laboratoire Leprince-Ringuet -  \'Ecole Polytechnique\\
        ${}^{b}$GRAPPA, Anton Pannekoek Institute for Astronomy, University of Amsterdam\\
        ${}^{c}$Erlangen Centre for Astroparticle Physics\\
        ${}^{d}$LPNHE Paris\\
        ${}^{e}$Max-Planck-Institut f\"ur Kernphysik\\
        ${}^{f}$Humboldt-Universit\"at zu Berlin\\
        ${}^{g}$Linnaeus University\\
        E-mail: \email{holler@llr.in2p3.fr}}

\abstract{The High Energy Stereoscopic System (H.E.S.S.) phase I instrument was an array of four $100\,\mathrm{m}^{2}$ mirror area Imaging Atmospheric Cherenkov Telescopes (IACTs) that has very successfully mapped the sky at photon energies above $\sim 100\,\mathrm{GeV}$. Recently, a $600\,\mathrm{m}^2$ telescope was added to the centre of the existing array, which can be operated either in standalone mode or jointly with the four smaller telescopes. The large telescope lowers the energy threshold for gamma-ray observations to several tens of GeV, making the array sensitive at energies where the \textit{Fermi}-LAT instrument runs out of statistics. At the same time, the new telescope makes the H.E.S.S. phase II instrument. This is the first hybrid IACT array, as it operates telescopes of different size (and hence different trigger rates) and different field of view.
In this contribution we present results of H.E.S.S. phase II observations of the Crab Nebula, compare them to earlier observations, and evaluate the performance of the new instrument with Monte Carlo simulations.}

\FullConference{The 34th International Cosmic Ray Conference,\\
		30 July- 6 August, 2015\\
		The Hague, The Netherlands}

\begin{document}

\newcommand{\hess}{H.E.S.S.}

\section{Introduction}
\label{introduction}

The High Energy Stereoscopic System (\hess ) played an important role in the rapid advancement of ground-based gamma-ray astronomy of the last dozen years. The greater part of all currently known very high energy (VHE; $E \gtrsim 100\,\mathrm{GeV}$) gamma-ray sources were discovered with the four Imaging Atmospheric Cherenkov Telescopes (IACTs) of \hess\ phase I, which are named CT1-4. These identical IACTs with an effective mirror diameter of $12\,\mathrm{m}$ are arranged on a square with edge length $120\,\mathrm{m}$ in the Khomas highland in Namibia. \hess\ phase II was initiated when the array was extended with a fifth telescope (CT5) with a $28\,\mathrm{m}$ mirror diameter in 2012, which is able to detect much fainter air showers. CT5 is located in the centre of the array and makes \hess\ the only IACT system that consists of telescopes with different sizes. Due to the greater complexity of such a hybrid array, advanced analysis and reconstruction methods are required to properly handle the data. Developing these methods can be considered as groundwork for the future Cherenkov Telescope Array (CTA) which will also be a hybrid system. 

Here we present results of the performance of the five-telescope \hess\ array. After introducing the analysis methods in Section~\ref{methods}, results of the Crab Nebula as a test source will be shown in Section~\ref{crab}. The performance for low-zenith angle observations will be evaluated in Section~\ref{performance}, followed by the conclusions in Section~\ref{conclusions}. 

\section{Analysis Methods}
\label{methods}

During \hess\ phase I, only stereoscopic triggers were read out to better suppress the hadronic background and improve the reconstruction performance (\cite{2004_HESS_Trigger}). With its roughly six times larger mirror area, CT5 is sensitive at energies way below the threshold of CT1-4. The data recorded with the \hess\ telescopes now consists of both single-telescope triggers from CT5 and stereoscopic triggers from CT1-5.

\begin{figure}
\center
\includegraphics[width = 0.7 \textwidth]{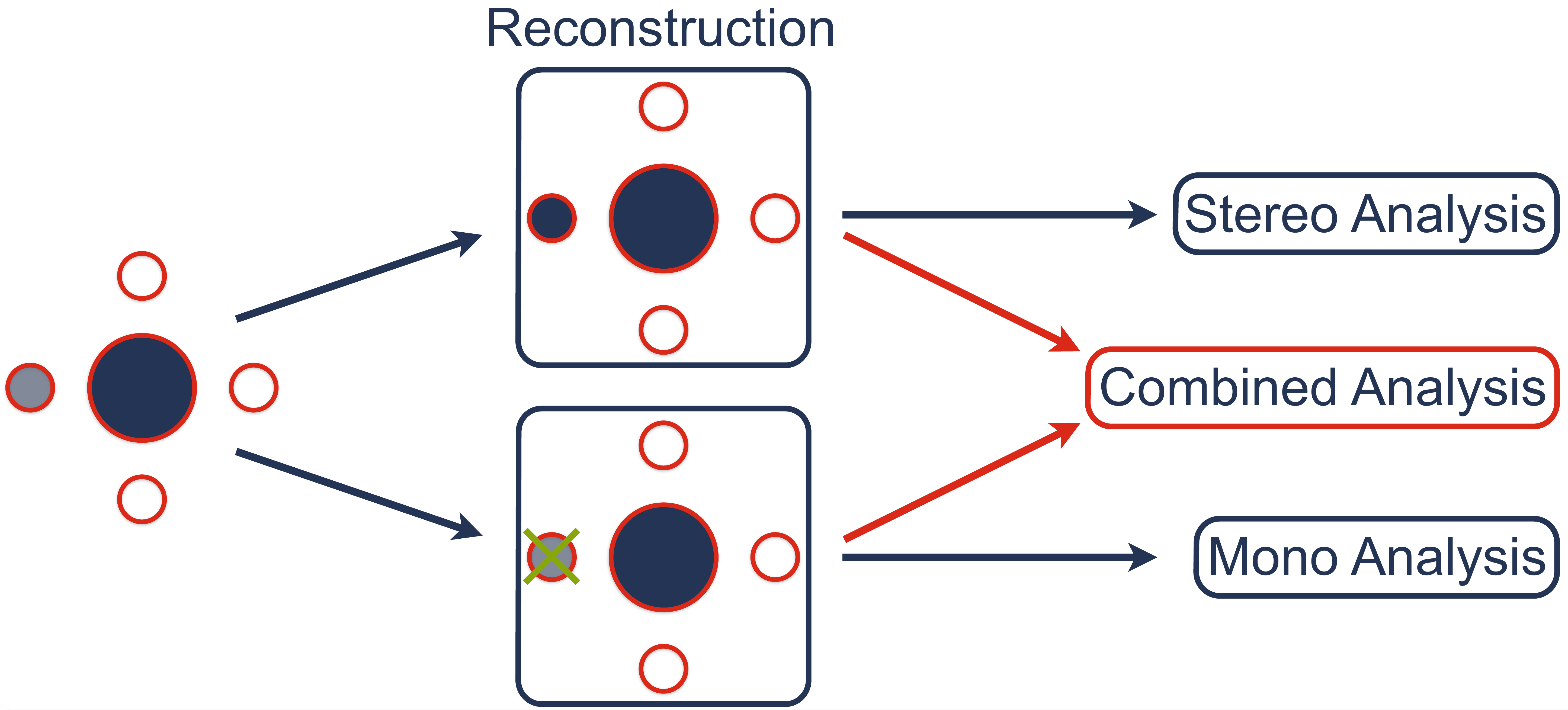}
\caption{A sketch of the different reconstruction and analysis modes (see text for more information).}
\label{sketch_modes}
\end{figure}
Fig.~\ref{sketch_modes} illustrates the different reconstruction and analysis modes of \hess\ II. Stereoscopic events where CT5 is part of the triggered telescopes are reconstructed both in single-telescope and stereoscopic mode. Source analyses can be performed with three different modes: the \textit{Stereo} analysis is similar to the one of \hess\ phase I but adapted to the five-telescope array. When applying the \textit{Mono} mode, only the information of CT5 is used, neglecting the other four telescopes of the array. To fully exploit all the available information, the \textit{Combined} mode makes use of both monoscopic and stereoscopic events, thus providing the best energy coverage. In case an event was reconstructed both monoscopically and stereoscopically, the estimate on the uncertainty of the reconstructed direction is used as a decision criterion which result to use (\cite{2015_Model}).

All results shown in the following were obtained with the \textit{Combined} analysis mode from \cite{2015_Model}. Cross-checks were performed with the single-telescope reconstruction method from \cite{2015_MonoReco} as well as a preliminary version of the ImPACT code (see \cite{2015_Impact}) that also combines the monoscopic and stereoscopic events. All analysis chains provide satisfactorily consistent results.

\section{Observations of the Crab Nebula}
\label{crab}

Being the brightest steady point source in the VHE gamma-ray sky, the Crab Nebula is an ideal test target for IACT arrays. Due to the location of H.E.S.S. in the Southern hemisphere, the Crab Nebula as a northern source is observed under zenith angles of at least $45^{\circ}$. We selected $7.47\,\mathrm{h}$ of good-quality observations with at least four participating telescopes, with CT5 always being one of them. The data consists of $16$ individual observation runs with a duration of $28\,\mathrm{min}$ each, with zenith angles ranging from $45^{\circ}$ to $55^{\circ}$. They were taken in wobble mode with offsets of $\pm 0.5^{\circ}$ in both right ascension and declination, respectively.

\begin{figure}
\center
\includegraphics[width = \textwidth]{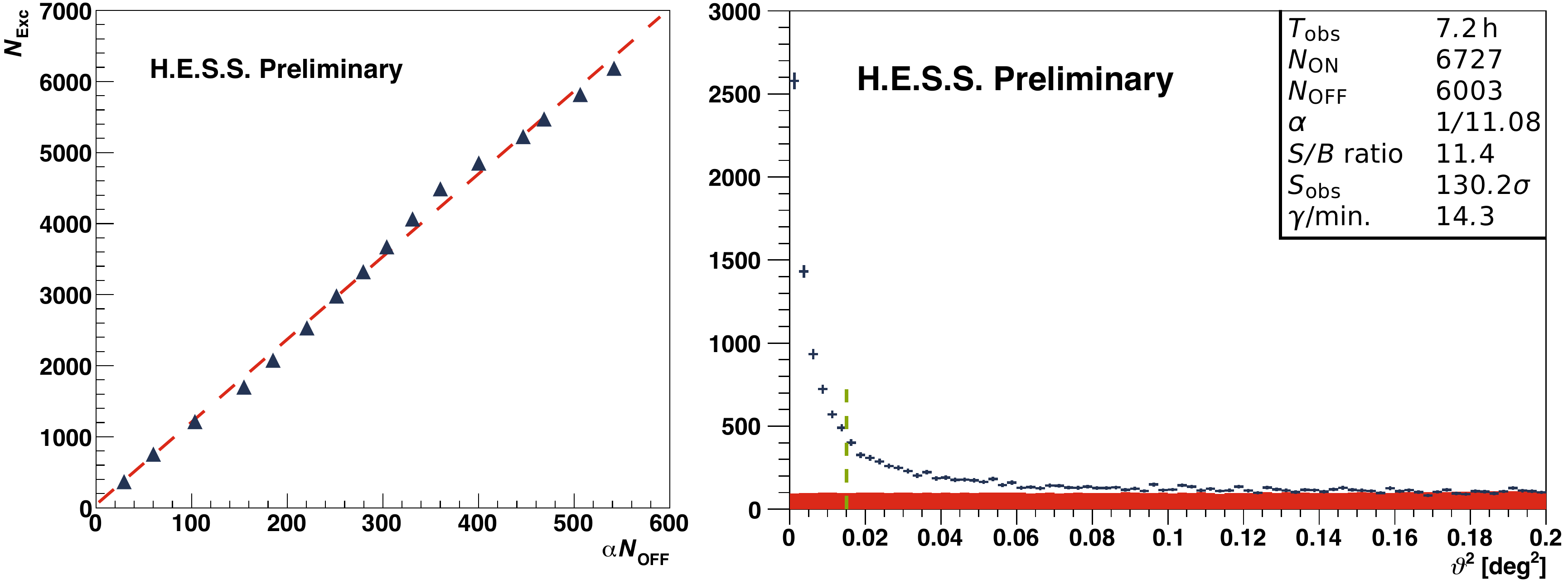}
\caption{\textit{Left:} evolution of the number of excess events as a function of the number of background events, the latter being scaled to be comparable to the source region. Each triangle denotes an observation run. \textit{Right:} angular distribution of events from the Crab Nebula after analysis cuts. The distribution around the source is shown in blue, and the one from the background regions (scaled with $\alpha$) in red. The $\vartheta^2$ cut, which was optimised for point-like sources analysed with the \textit{Mono} mode (see \cite{2015_Model}), is indicated with a dashed green line. The event statistics are shown on the top right.}
\label{theta2}
\end{figure}
To illustrate the data quality at analysis level, the evolution of the excess with respect to the background level is shown in the \textit{left panel} of Fig.~\ref{theta2}, calculated using the \textit{reflected-region} method from \cite{2007_Berge_Background}. For similar observation conditions and a steady source it is expected to rise linearly, with an additional scattering that is introduced by statistical and systematic errors. The apparent steady progression indicates a constant data quality. The $\vartheta^2$ histogram of gamma-like events of the whole data set is shown in the \textit{right panel} of Fig.~\ref{theta2}, together with the event statistics on the top right. The angular distributions of source and background are well normalised at larger offsets. The source is detected at a large significance and with a very high $S/B$ ratio. It has to be noted that, according to Monte-Carlo (MC) simulations, the expected excess rate would be even $\approx 40\%$ higher if the source could be observed at low zenith angles.

\begin{figure}
\center
\includegraphics[width = \textwidth]{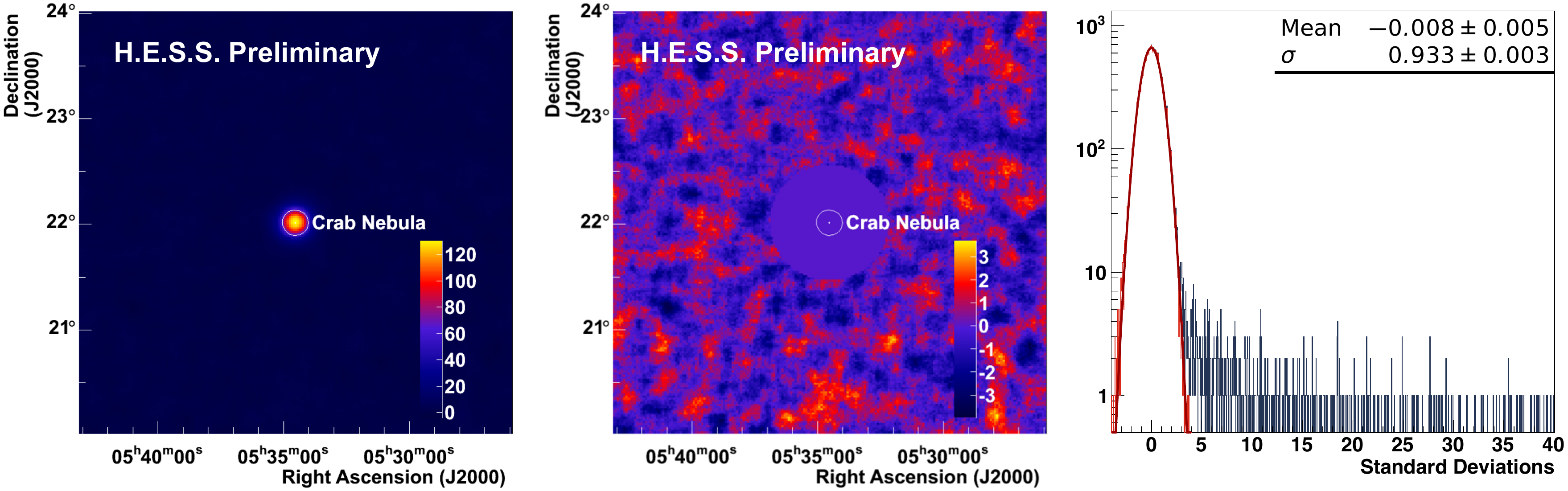}
\caption{\textit{Left:} significance sky map of the Crab Nebula as seen with \hess\ II in \textit{Combined} mode. \textit{Middle:} same map as in the \textit{left panel}, but with a region around the target being blinded to enhance the visibility of background fluctuations. \textit{Right:} projection of values of the map, only showing values up to $40\sigma$ for visibility reasons. The blue histogram contains all entries of the map, whereas the red one corresponds to all bins outside the target region. The latter was fitted with a Gaussian whose fit results are shown on the top right.}
\label{sig_map}
\end{figure}
Fig.~\ref{sig_map} (\textit{left panel}) contains the significance map of the region, calculated using the \textit{ring}-background method from \cite{2007_Berge_Background}. The Crab Nebula clearly appears at the expected position as a point-like source (though detailed comparisons of the simulated and measured angular resolution still have to be carried out). To better check for potential issues outside the target region, the \textit{middle panel} of Fig.~\ref{sig_map} shows the same map but with a region around the source being blinded. There are no significant fluctuations apparent, and the map is well normalised (see \textit{right panel} of Fig.~\ref{sig_map}). 

During \hess\ phase I, the spectral shape of the Crab Nebula was sufficiently described with a power law with an exponential cut-off (see \cite{2006_Hess_Crab}). Since MAGIC and Veritas can observe the source at lower zenith angles and thus with a lower energy threshold, a log-parabola model is needed to properly model the spectral curvature over the whole energy range (\cite{2015_MAGIC_Spectrum},\cite{2015_CrabVeritas}).
\begin{figure}
\center
\includegraphics[width = 0.7 \textwidth]{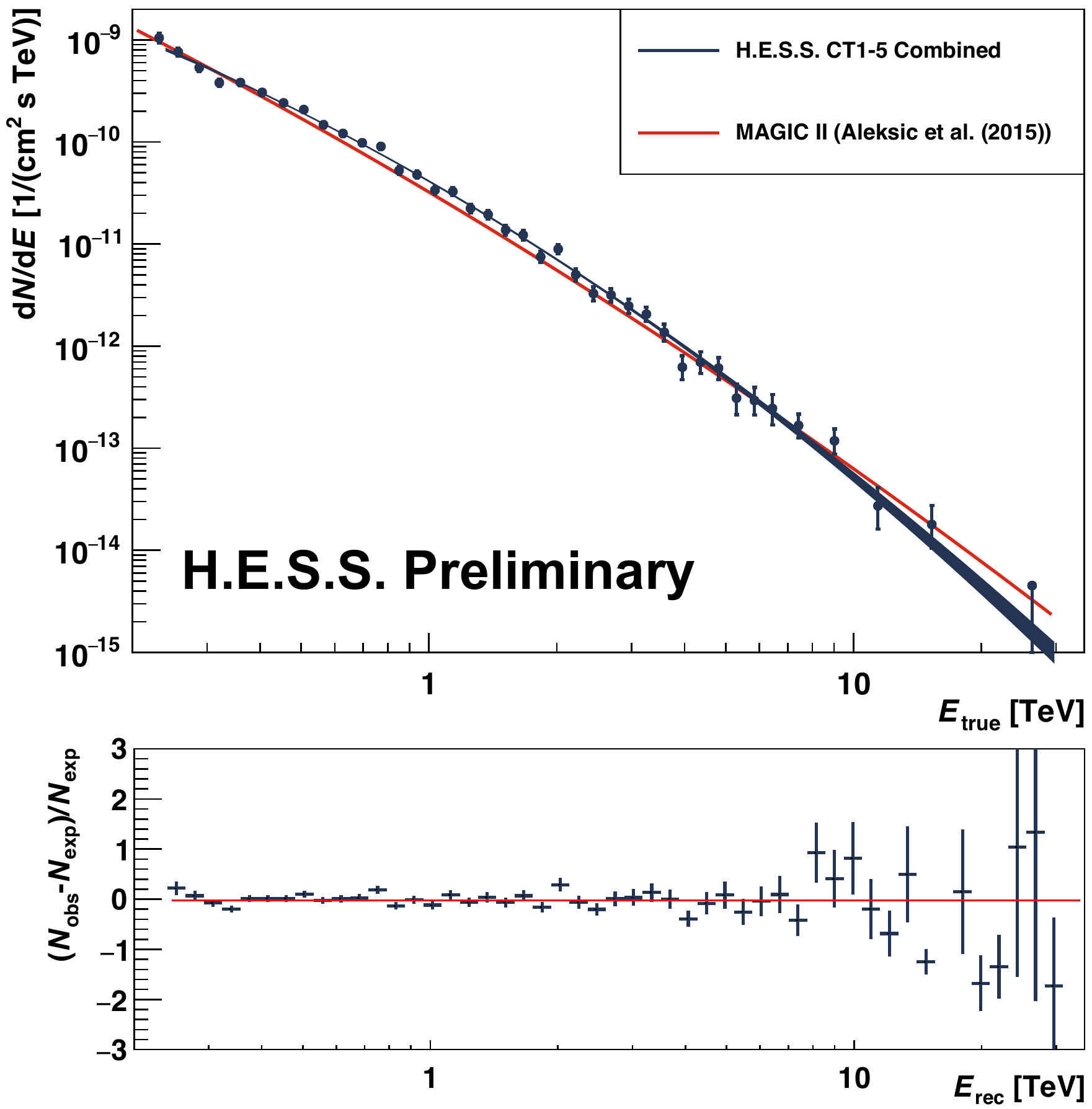}
\caption{Energy spectrum of the Crab Nebula, measured with \hess\ II using the \textit{Combined} mode. The comparison spectrum is the one from \cite{2015_MAGIC_Spectrum}. The fit residuals are drawn in the bottom.}
\label{spectrum}
\end{figure}
The differential energy spectrum of the Crab Nebula as measured with \hess\ II using the \textit{Combined} analysis mode is shown in Fig.~\ref{spectrum}. The energy threshold is $\approx 230\,\mathrm{GeV}$ as compared to $\approx 440\,\mathrm{GeV}$ with \hess\ I (\cite{2006_Hess_Crab}). The spectrum is best described by a log-parabola function as follows:
\begin{equation}
\label{eq_spectrum}
\frac{\mathrm{d} N}{\mathrm{d} E} = \left( 1.79 \pm 0.03 \right)\times 10^{-10}~ \left( \frac{E}{0.521\mathrm{TeV}} \right)^{-(2.10\pm 0.04)-(0.24\pm 0.01)\, \cdot\, \ln(\frac{E}{0.521\mathrm{TeV}})} \frac{1}{\mathrm{TeV}\, \mathrm{cm}^{2}\, \mathrm{s}}~\mathrm{.}
\end{equation}
All errors stated in Eq.~\ref{eq_spectrum} are purely statistical ones. An investigation of systematic uncertainties will be part of a future publication.

\section{Performance}
\label{performance}

To assess the performance of \hess\ II for observations at low zenith angles, MC simulations of gamma-rays were generated. More precisely, we simulated a point source at $18^{\circ}$ zenith angle, $180^{\circ}$ azimuth angle (i.e. with the telescopes pointing to the south), and $0.5^{\circ}$ wobble offset.

\begin{figure}
\center
\includegraphics[width = 0.8 \textwidth]{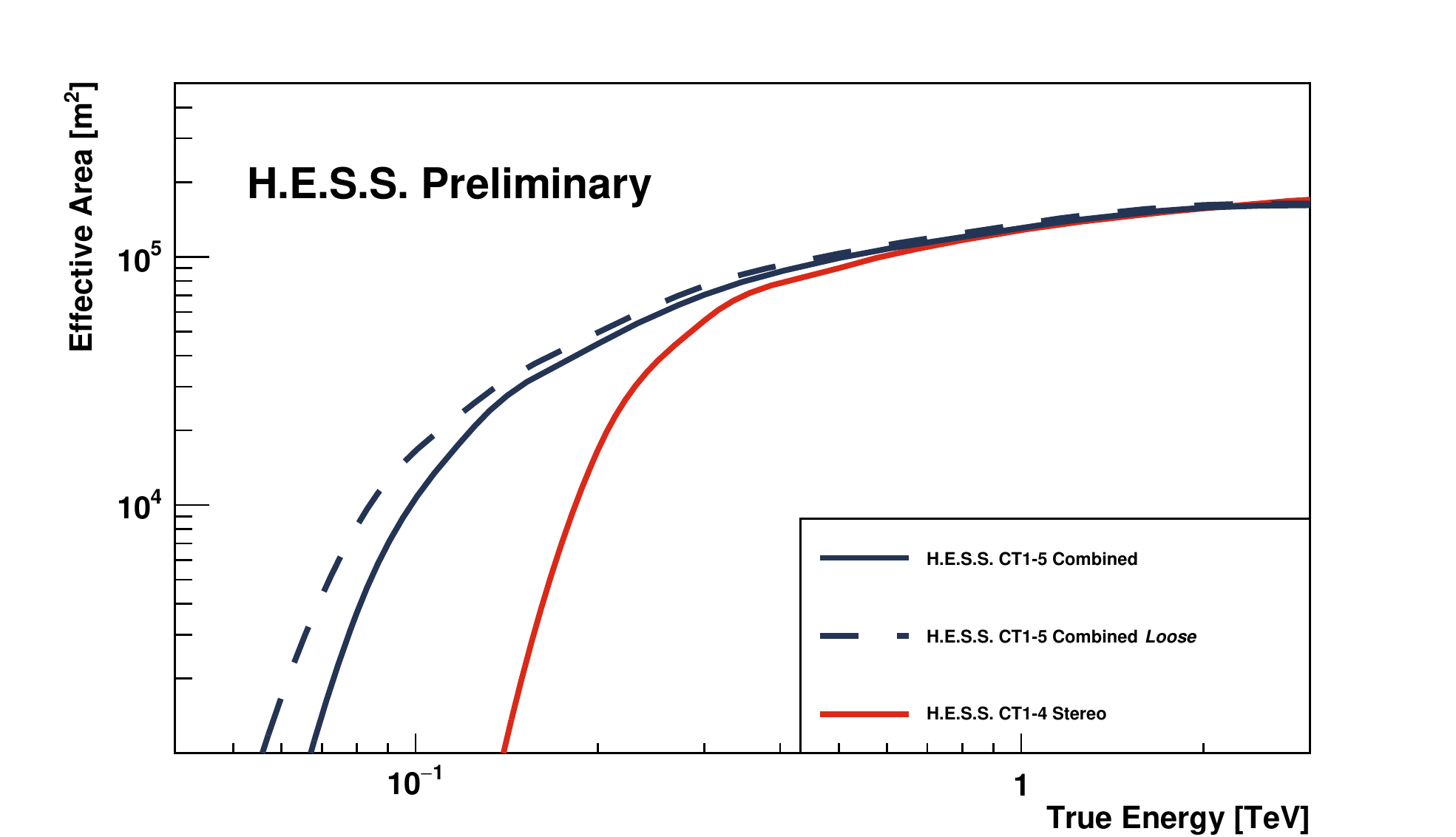}
\caption{Analysis effective area for low-zenith angle observations. The solid blue line corresponds to the \textit{Standard} cut configuration that is defined in \cite{2015_Model}, and the dashed curve to a looser configuration.}
\label{spectrum}
\end{figure}
The analysis effective area is shown in Fig.~\ref{spectrum}. As for \hess\ II in \textit{Combined} mode, curves for two different cut sets are drawn: the solid blue line corresponds to the \textit{Standard} cut configuration that is defined in \cite{2015_Model} and which is applicable to almost all sources, and the dashed curve to a looser configuration which provides a lower energy threshold. The latter was used to obtain the results in Section~\ref{crab}. At $E \gtrsim 300\,\mathrm{GeV}$, the effective area of \hess\ II connects with the one of \hess\ I, indicating that the addition of CT5 mostly influences the array at energies below. With the current conservative analysis cuts, it is possible to reconstruct spectra down to energies of $\approx 70-80\,\mathrm{GeV}$ (\cite{2015_HESS_AGN}).

To evaluate the overall source detection performance of \hess\ II, the differential sensitivity was calculated. Background events were taken from actual data with observation zenith angles between $12^{\circ}$ and $22^{\circ}$ to closely match the zenith angle of the simulations.
\begin{figure}
\center
\includegraphics[width = \textwidth]{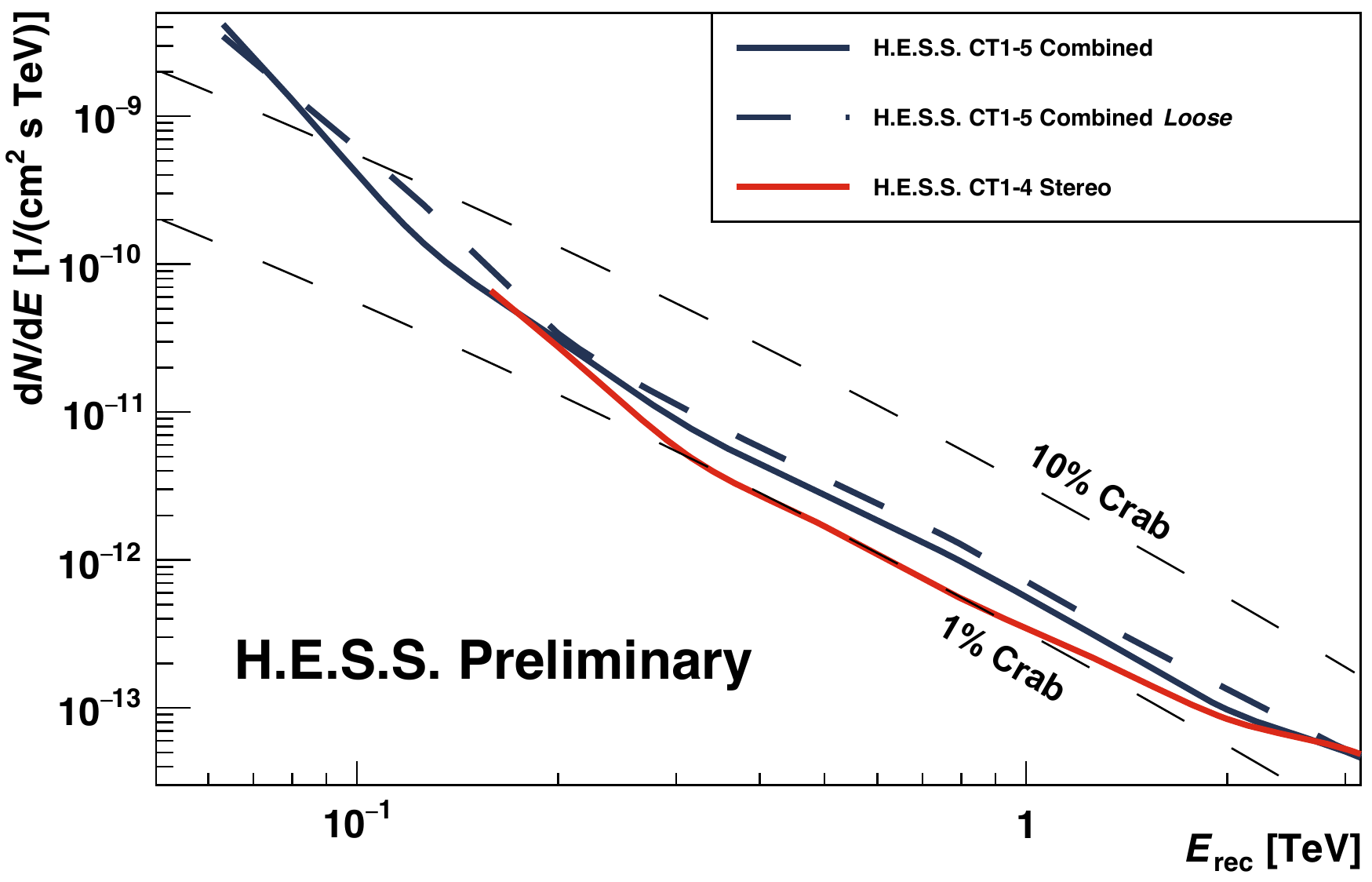}
\caption{Differential sensitivity of \hess\ II using the \textit{Combined} analysis mode as compared to the one of the original four-telescope array. The reference spectrum used for the dashed black lines is the one from \cite{2015_MAGIC_Spectrum}.}
\label{diff_sens}
\end{figure}
Fig.~\ref{diff_sens} shows the minimum source strength that is required to gain a detection of $5\sigma$ after an observation time of $50\,\mathrm{h}$ as a function energy. Each of the curves consists of $5$ bins per decade and was created using a simplified significance calculation $N_{\gamma}/\sqrt{N_{\textrm{bkg}}}$. For each bin, a minimum of $10$ signal events as well as $S/B \ge 0.05$ was requested. As expected, CT5 makes the array sensitive at energies too low to be accessible by CT1-4. In turn, the \textit{Combined} analysis in its current form is not yet as sensitive as the \textit{Stereo} analysis at medium and high energies, which can be mostly be accounted to the fact that the selection cuts still have to be optimised for performance across the full energy range (see \cite{2015_Model}). The comparably large $\vartheta^2$ cut (which is currently the same as for the \textit{Mono} mode) for instance implies a considerably higher background level.

\section{Conclusions}
\label{conclusions}

We successfully presented results from the full hybrid \hess\ array, applying a method which combines monoscopic and stereoscopic events into one overall analysis. 

Observations of the Crab Nebula were used to test the method. All results show that \hess\ II performs well in terms of data-taking, calibration, reconstruction, and analysis. Despite the large zenith angle, the source is detected at a high significance rate of $48\sigma/\sqrt{\mathrm{h}}$. Thanks to CT5, the energy threshold lies almost $50\%$ below the one of \hess\ I. 

To complement the results of the Crab Nebula, the performance of low-zenith angle observations was evaluated with simulated gamma-rays and real background events. Adding CT5 to the array led to a considerable decrease of the energy threshold down to currently $\approx 80\,\mathrm{GeV}$. In case systematic uncertainties can be reduced (like for pulsar observations), it is possible to loosen the analysis cuts to even detect gamma-rays with $E \lesssim 20\,\mathrm{GeV}$ (\cite{2015_Vela}). A better assessment and control of these uncertainties will allow to also lower the threshold for standard (i.e. non-pulsed) analyses.

\section{Acknowledgements}

The support of the Namibian authorities and of the University of Namibia in facilitating the construction and operation of H.E.S.S. is gratefully acknowledged, as is the support by the German Ministry for Education and Research (BMBF), the Max Planck Society, the German Research Foundation (DFG), the French Ministry for Research, the CNRS-IN2P3 and the Astroparticle Interdisciplinary Programme of the CNRS, the U.K. Science and Technology Facilities Council (STFC), the IPNP of the Charles University, the Czech Science Foundation, the Polish Ministry of Science and Higher Education, the South African Department of Science and Technology and National Research Foundation, and by the University of Namibia. We appreciate the excellent work of the technical support staff in Berlin, Durham, Hamburg, Heidelberg, Palaiseau, Paris, Saclay, and in Namibia in the construction and operation of the equipment.

\bibliographystyle{JHEP}
\bibliography{crab_proceeding}

\providecommand{\href}[2]{#2}\begingroup\raggedright\begin{thebibliography}{10}

\bibitem{2004_HESS_Trigger}
S.~{Funk}, G.~{Hermann}, J.~{Hinton}, D.~{Berge}, K.~{Bernl{\"o}hr},
  W.~{Hofmann}, P.~{Nayman}, F.~{Toussenel}, and P.~{Vincent}, {\it {The
  trigger system of the H.E.S.S. telescope array}},  {\em Astroparticle
  Physics} {\bf 22} (2004) 285--296,
  [\href{http://arxiv.org/abs/astro-ph/0408375}{{\tt astro-ph/0408375}}].

\bibitem{2015_Model}
M.~{Holler et al.}, {\it Photon reconstruction for H.E.S.S. using a
  semi-analytical shower model},  {\em Proceedings of the 34th International
  Cosmic Ray Conference} (2015).

\bibitem{2015_MonoReco}
T.~{Murach et al.}, {\it A Neural Network-Based Monoscopic Reconstruction
  Algorithm for H.E.S.S. II},  {\em Proceedings of the 34th International
  Cosmic Ray Conference} (2015).

\bibitem{2015_Impact}
D.~{Parsons et al.}, {\it H.E.S.S. II Data Analysis with ImPACT},  {\em
  Proceedings of the 34th International Cosmic Ray Conference} (2015).

\bibitem{2007_Berge_Background}
D.~{Berge et al.}, {\it Background modelling in very-high-energy gamma-ray
  astronomy},  {\em A\&A} {\bf 466} (2007) 1219--1229,
  [\href{http://arxiv.org/abs/astro-ph/0610959}{{\tt astro-ph/0610959}}].

\bibitem{2006_Hess_Crab}
F.~{Aharonian et al.}, {\it Observations of the Crab nebula with HESS},  {\em
  A\&A} {\bf 457} (2006) 899--915,
  [\href{http://arxiv.org/abs/astro-ph/0607333}{{\tt astro-ph/0607333}}].

\bibitem{2015_MAGIC_Spectrum}
J.~{Aleksi\'c et al.}, {\it Measurement of the Crab Nebula spectrum over three
  decades in energy with the MAGIC telescopes},  {\em Journal of High Energy
  Astrophysics} {\bf 5 to 6} (2015) 30 -- 38.

\bibitem{2015_CrabVeritas}
K.~{Meagher et al.}, {\it Six years of VERITAS observations of the Crab
  Nebula},  {\em Proceedings of the 34th International Cosmic Ray Conference}
  (2015).

\bibitem{2015_HESS_AGN}
D.~{Zaborov et al.}, {\it AGN observations with a less than 100 GeV threshold
  using H.E.S.S. II},  {\em Proceedings of the 34th International Cosmic Ray
  Conference} (2015).

\bibitem{2015_Vela}
M.~{Gajdus et al.}, {\it Pulsations from the Vela Pulsar down to 20 GeV with
  H.E.S.S. II},  {\em Proceedings of the 34th International Cosmic Ray
  Conference} (2015).

\end{thebibliography}\endgroup

\end{document}